\documentclass[aps,showpacs,twocolumn]{revtex4}
\usepackage[dvips]{graphics,graphicx}
\usepackage{amsmath}
\usepackage{amssymb}                 
\usepackage{amscd}                   
\usepackage{wasysym}
\newcommand{\ri}{{ \rm i }}
\newcommand{\re}{{ \rm e }}
\newcommand{\rd}{{ \rm d }}

\usepackage[ansinew]{inputenc}
\usepackage[T1]{fontenc}
\usepackage{ae,aecompl}

\newcommand{\be}{\begin{equation}}
\newcommand{\ee}{\end{equation}}

\newcommand{\leapprox}{\apprle}

\DeclareGraphicsExtensions{.eps}
\begin{document}
\bibliographystyle{apsrev}
\title{Bloch oscillations of Bose-Einstein condensates: Breakdown and revival}
\author{D. Witthaut, M. Werder, S. Mossmann, and H. J. Korsch}
\email{korsch@physik.uni-kl.de}
\affiliation{Technische Universit{\"a}t Kaiserslautern, FB Physik,
                            D-67653 Kaiserslautern, Germany}
\date{\today }

\begin{abstract}
We investigate the dynamics of Bose--Einstein condensates (BEC)
in a tilted one--dimensional periodic lattice within the mean--field (Gross--Pitaevskii)
description. Unlike in the linear case the Bloch oscillations decay because of
nonlinear dephasing. Pronounced revival phenomena are observed. These are
analyzed in detail in terms of a simple integrable model
constructed by an expansion in Wannier--Stark resonance states.
We also briefly discuss the pulsed output of such systems for stronger static fields.
\end{abstract}

\maketitle


\section{Introduction}
Despite its apparent simplicity, the dynamics of quantum particles in periodic structures is full of surprises, even in the one--dimensional case.
For almost a century Bloch waves have been known which are delocalized states in a lattice
leading to transport.
If an additional static field $F$ is introduced, these states become localized
and counterintuitively transport is dramatically reduced. Instead
an oscillatory motion is found, the famous Bloch oscillations. These have a (Bloch)
frequency $\omega_B=Fd/\hbar$ where $d$ is the lattice constant and they extend
over a spatial interval $\Delta /F$ where $\Delta$ is the width of the first Bloch
band. During the last decade these Bloch oscillations have been experimentally
observed which triggered a renewed theoretical interest (for recent
reviews see \cite{04bloch1d,03bloch2D,04bloch2d}).

For stronger fields decay has to be taken into account. So
this simple picture must be replaced by introducing
couplings to higher bands or, alternatively, by a description in terms of
Wannier--Stark resonances. More details can be found in the review article
\cite{02wsrep}.

One of the most interesting systems for exploring the dynamics described above
are cold atoms in optical lattices, because here the notorious difficulties
met in solid state systems (where, in fact, the Bloch oscillations have been
observed for the first time \cite{Feld92}) are absent or, at least, can be made
very small. Kasevich did one of the first experiments
\cite{Ande98} with a Bose--Einstein condensate (BEC) of Rubidium atoms in a
optical lattice with gravity acting as the static field. They could observe
a pulsed coherent output of atoms.

The atoms in a BEC scatter off each other,
which offers the opportunity to study the influence of the atomic
interaction on the dynamics.
In a good approximation, the dynamics can be described by the
one--dimensional Gross--Pitaevskii equation (GPE) \cite{Chio00}
\begin{eqnarray}
\label{GPE}
{\rm i}\hbar\, \partial_t \psi = \Big[-\frac{\hbar^2}{2M}\,\partial^2_x  + V(x) +
Fx + g \,|\psi|^2\,\Big] \,\psi\, ,
\end{eqnarray}
where $M$ is the atomic mass, $g$ is the interaction strength.
and \,$V(x)=V(x+d)$\, is the periodic lattice potential.

This nonlinear system shows basically all the features found in the analysis
of the linear equation, such as Bloch oscillations of the condensate
\cite{Berg98,Choi99}.
In addition, the nonlinearity introduces new effects, such as soliton--like
motion, nonlinear Zener tunneling \cite{Wu00, Wu03}  and "classically"
chaotic dynamics \cite{Trom01, Thom03, Kolo04}.
This system has been analyzed with various methods, see
e.g. \cite{Holt00,Java99,Thom04}.

In the present article, we will focus on Bloch oscillations and analyze
the influence of the nonlinearity. The paper is organized as follows:
In section \ref{sec-numerical} we present results from a numerical solution
of the GPE and show
nonlinear Bloch oscillations for relatively weak fields and different
strength and signs of the nonlinear interaction. Section
\ref{sec-WS-expansion} introduces our main tool, a discrete
representation by an expansion in Wannier--Stark resonance states and
derives approximate results based on this approach.
These results are used to analyze the dynamical behavior of Bloch
oscillations in section \ref{sec-koeffizienten-dynamik}.
Finally we discuss the modification
of the coherent pulse output of a Bloch oscillating condensate
for stronger fields in section \ref{sec-basis-hohe-felder}. The
paper closes with some concluding remarks.
\section{Numerical study of nonlinear Bloch oscillations}
\label{sec-numerical}

Due to the nonlinearity of the GPE (\ref{GPE}) analytical studies are difficult,
so numerical simulations are helpful in guiding theoretical investigations.

In all numerical studies we
will use a cosine potential \,$V(x)=V_0\cos (2\pi x/d)$.
We furthermore use scaled
units with $d=2\pi$, $V_0 = M = 1$ (see \cite{02wsrep} for more
details). It is worth noting that the scaled interaction strength is inversely
proportional to the depth of the potential.

The GPE then reads
\begin{eqnarray}
\label{boseScUn}
  {\rm i}\hbar\, \partial_t \psi = \Big[-\frac{\hbar^2}{2}\,\partial^2_x
  + \cos x + Fx + g \,|\psi|^2\,\Big] \,\psi
\end{eqnarray}
and we will use the value $\hbar = 3.3806$ for the
scaled Planck constant adapted to the experiment of the
Kasevich group \cite{Ande98} (see also \cite{02pulse}).
The nonlinearity parameter is of the order $g \leapprox 1$ in this
experiment. Here we will extend the analysis, however, to much
stronger nonlinearities up to $|g| =10$. This regime could be reached
experimentally by increasing the transverse confinement or decreasing
the depth of the optical lattice.

Here we are mainly interested in the dynamics of Bloch oscillations
and therefore use a weak field ($F = 0.005$) and initial states
populating almost exclusively the lowest Bloch band. In this case
the decay is negligible.
In the linear case the band gap between the lowest and the next
higher Bloch band for the field free case is $\delta = 0.998$
and the probability for Zener tunneling is $\approx 10^{-12}$.
This does not change noticeably in the nonlinear case as has been
checked numerically, showing that nonlinear Zener tunneling \cite{Wu00, Wu03}
does not play a role for the given parameters.
Furthermore note that the dynamical effects which are observed in the
following examples (e.g. the damping of Bloch oscillations)
are essentially captured by a simple integrable model
which is introduced in section \ref{sec-WS-expansion}.
This should be compared to the breakdown of Bloch oscillations due
to dynamical instability discussed in \cite{Wu03,Liu03,Meno03}.

\begin{figure}[htb]
\centering
\includegraphics[width=7cm,  angle=0]{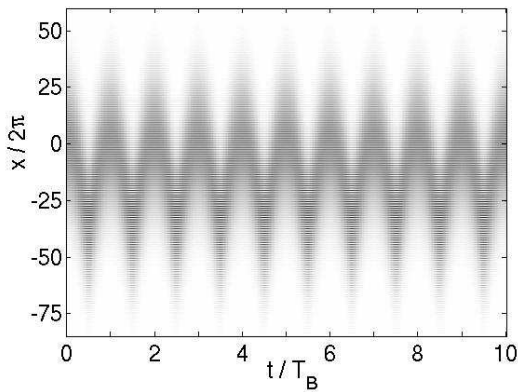}
\caption{\label{fig-bonl_g=0}
The squared modulus $|\psi(x,t)|^2$ of the wave function
for $g=0$ shows the familiar Bloch oscillations.}
\end{figure}
\begin{figure}[htb]
\centering
\includegraphics[width=8cm,  angle=0]{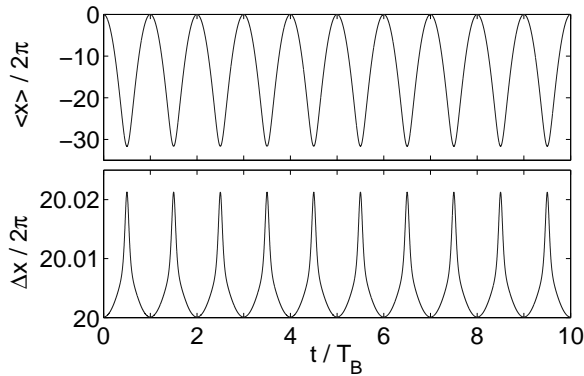}
\caption{\label{fig-erw_g=0}
Expectation values of the position $\langle x \rangle_t$ and
width $\Delta x_t$ for the wave function shown in figure
\ref{fig-bonl_g=0}.}
\end{figure}

Let us start our discussion with a brief look at the Bloch
oscillation for the linear case $g=0$. As an initial state,
we use a Gaussian wave packet
\be
  \psi(x,t=0) = \frac{1}{(2 \pi)^{1/4} \tilde\sigma^{1/2}}
  \re^{-(x-x_0)^2/4\tilde\sigma^2}
\ee
with width $\tilde\sigma = 40 \pi$ which is then projected onto
the lowest Bloch band and normalized to unity. For the time propagation
a split--operator method \cite{Feit82}  is used which can also be
applied to the nonlinear case. In figure \ref{fig-bonl_g=0}
we observe the familiar Bloch oscillation with a large amplitude
because of the weak field.
Let us recall that the region over which the Bloch
oscillation extends can be estimated as
$\Delta /F\approx 200\approx 32\cdot 2\pi$ within the tilted band picture,
where $\Delta = 0.9994 $ is the width of the dispersion relation
$E(\kappa)$ in the field free case.
The numerical results confirm this estimate as the top of figure \ref{fig-erw_g=0} shows.
As expected for such an initially wide distribution in coordinate space,
the width of the wave packet remains practically constant varying
periodically with a relative amplitude of about $10^{-3}$ (figure \ref{fig-erw_g=0} bottom).

\begin{figure}[htb]
\centering
\includegraphics[width=7cm,  angle=0]{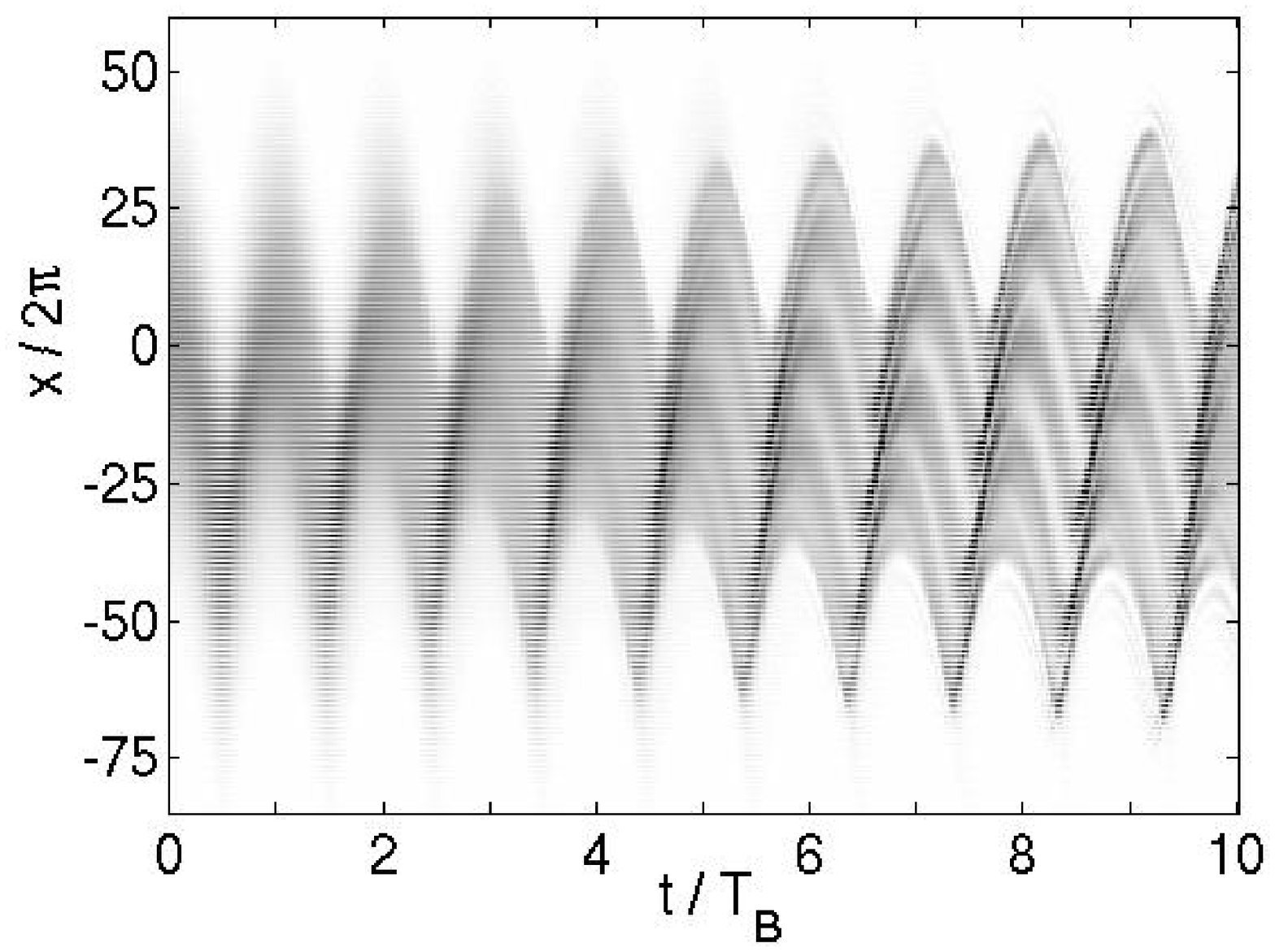}
\includegraphics[width=7cm,  angle=0]{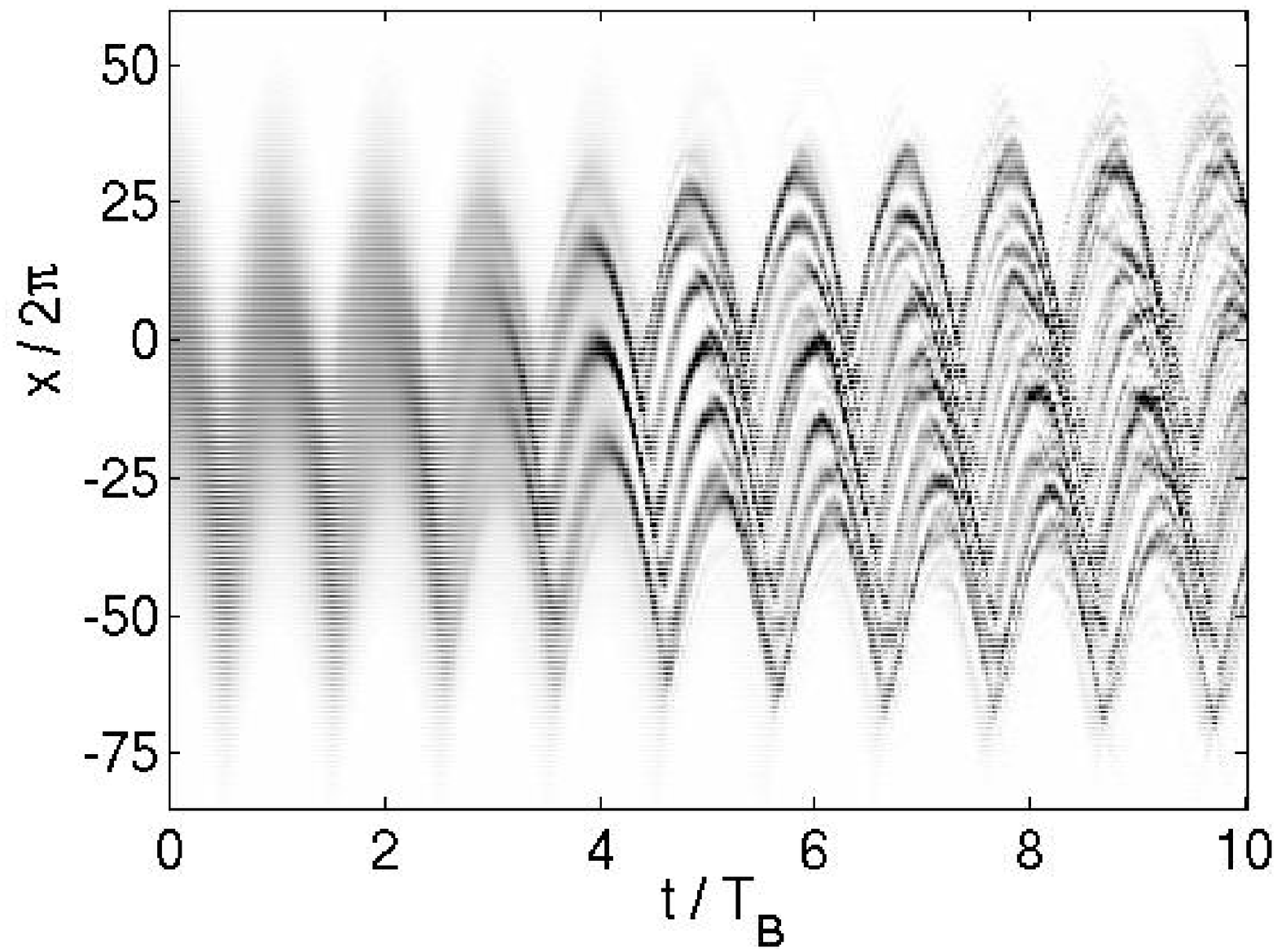}
\caption{\label{fig-bonl_g=pm5}
Same as figure \ref{fig-bonl_g=0}, however for a nonlinear
interaction  $g=+5$ (top) and $g=-5$ (bottom).}
\end{figure}

\begin{figure}[htb]
\centering
\includegraphics[width=8cm,  angle=0]{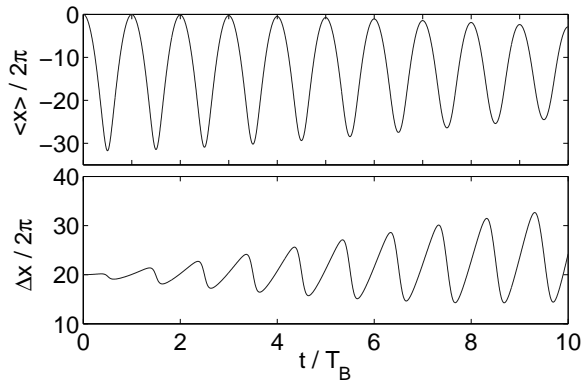}
\caption{\label{fig-bonl_xdx_g=p5}
Expectation values of the position $\langle x \rangle_t$ and
width $\Delta x_t$ for the wave function shown in figure
\ref{fig-bonl_g=pm5} for a repulsive nonlinearity $g=+5$.
}
\end{figure}

Let us now discuss the influence of a nonlinearity fixing for the
moment the nonlinear parameter at $g = 5$ (repulsive interaction)
and $g = -5$ (attractive interaction).
From figure \ref{fig-bonl_g=pm5} one can observe that the
the Bloch oscillations continue to exist, at least for the
short times up to $t \approx 10 \, T_B$ shown in the figure.
In addition to the well known localization
of $|\psi(x,t)|^2$ in the minima of the cosine--potential,
one observes a further filamentation which is particularly
pronounced for an attractive interaction.
As shown in figure \ref{fig-bonl_xdx_g=p5}, the amplitude
of the oscillation decreases and the oscillation
in the width strongly increases.

\begin{figure}[htb]
\centering
\includegraphics[width=8cm,  angle=0]{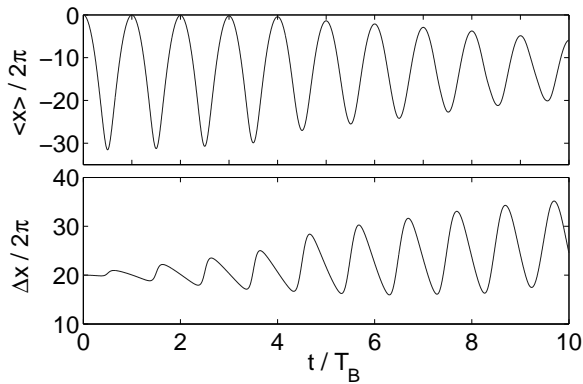}
\caption{\label{fig-bonl_xdx_g=m5}
Same as figure \ref{fig-bonl_xdx_g=p5}, however for
for an attractive nonlinearity $g=-5$.
}
\end{figure}
Also shown in figures \ref{fig-bonl_xdx_g=p5} and \ref{fig-bonl_xdx_g=m5}
is the time dependence of the width $\Delta x_t$  of the wave packet.
In sharp contrast to the tiny oscillations of the width in the linear
case (see figure \ref{fig-erw_g=0}) we find here very pronounced
oscillations which are rapidly growing (as already described by Holthaus
\cite{Holt00}). Such a phenomenon is known as {\it breathing}
and is exhibited in the linear system by wave packets which are
initially strongly localized in coordinate space \cite{03TBalg,04bloch1d}.
Note that the oscillations of the width $\Delta x_t$ for a repulsive
and attractive nonlinearity are opposite to each other.

\begin{figure}[htb]
\centering
\includegraphics[width=8cm,  angle=0]{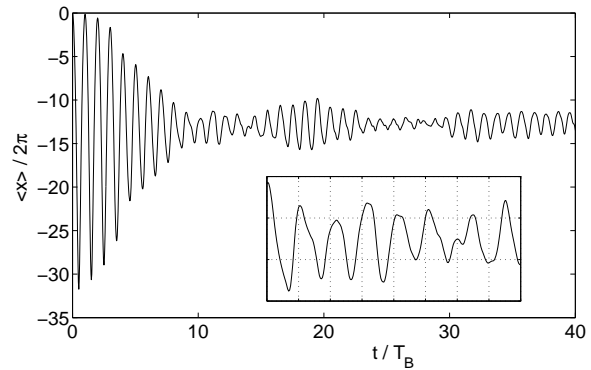}
\caption{\label{fig-bonl_long_xev}
Expectation value $\langle x \rangle_t$ of the position and
width $\Delta x_t$ for a repulsive nonlinearity $g=+10$.
The inset shows a magnification of the time interval between $8$ and $16\,T_B$.}
\end{figure}

\begin{figure}[htb]
\centering
\includegraphics[width=8cm,  angle=0]{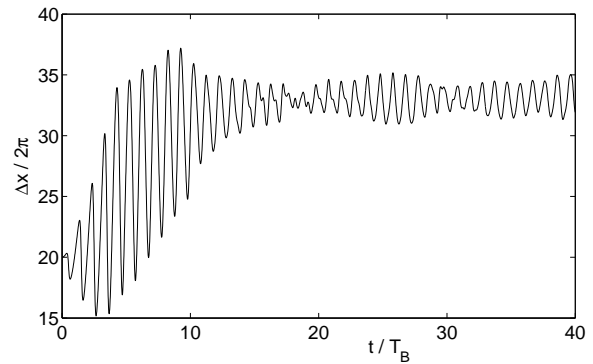}
\caption{\label{fig-bonl_long_dxev}
Width $\Delta x_t$ of the wave packet shown in figure \ref{fig-bonl_long_xev}
for a repulsive nonlinearity $g=+10$.}
\end{figure}

For a stronger nonlinearity $g=10$, as illustrated in figures \ref{fig-bonl_long_xev}
and \ref{fig-bonl_long_dxev}, the Bloch oscillations are damped more strongly.
However, the oscillation does not fade completely but shows
a revival with a smaller amplitude after a shrinking to approximately
two lattice periods. A corresponding behavior is observed for the width, where
the breathing amplitude of the wave function first grows fast up to a time of
about eight Bloch periods. After this time, the width remains limited and oscillates
in the interval from $31$ to $35$ lattice periods.

Furthermore, the oscillation of $\langle x \rangle_t$ shows phase
jumps that can be seen, e.g. in the inset of figure \ref{fig-bonl_long_xev} at
$t \approx 14 \, T_B$. A similar behavior has also been described in \cite{Holt00}.
This phase jump coincides with a minimum in the amplitude.
These phenomena can be understood in terms of an expansion in
Wannier-Stark basis functions as explained in the next section.

\section{Wannier--Stark basis set expansion}
\label{sec-WS-expansion}
An alternative approach to a direct numerical integration of the GPE is
an expansion in an adequate discrete basis as for example
the ground states of single potential wells \cite{Trom01,Abdu01}.
In this work we adopt a different approach,
following \cite{Thom03,Thom04},  and expand the wave function
into the resonance eigenstates of the
linear system, the so--called Wannier--Stark states $\Psi_{\alpha,n}(x)$
which are eigenstates of the linear Hamiltonian $H_0$:
\be
 H_0  \Psi_{\alpha,n}(x)
  = {\cal E}_{\alpha,n}  \Psi_{\alpha,n}(x)\, ,
\ee
where  $\alpha$ is the ladder index and $n$ is the
site index.
The energies form the Wannier--Stark ladder
\be
  {\cal E}_{\alpha,n} = {\cal E}_{\alpha,0} + 2 \pi F n.
\ee
The Wannier--Stark states extend over several periods of the
potential (see remark \cite{remark04} and review \cite{02wsrep} for more information).
This approach has proven to be extremely convenient to describe the dynamics
in tilted optical lattices in the linear case, especially for higher
field strengths \cite{02wsrep,02pulse}.

Up to section \ref{sec-basis-hohe-felder}, we will restrict the discussion
to small field strengths $F$. Then one can neglect decay and Landau--Zener
tunneling and use the lowest ladder $\alpha = 0$ only, henceforth the
index $\alpha$ is omitted.
Also neglecting decay, the imaginary part of the energy ${\cal E}_{n}$
is set to zero.
Plugging the expansion $\psi(x,t) = \sum_m c_m(t) \Psi_m(x)$ into the GPE (\ref{GPE})
leads to a set of coupled ordinary differential equations:
\begin{eqnarray}
  \ri  \hbar \sum_m \dot c_m \Psi_m &=& \sum_m ({\cal E}_0 + 2 \pi F m)
  \, c_m \Psi_m \nonumber \\
  && \qquad + g \sum_{klm} c_k^* c_l c_m \Psi_k^* \Psi_l \Psi_m.
  \label{eqn-nlse-wsbasis}
\end{eqnarray}
The energy ${\cal E}_0$ only leads to a global phase factor and
hence is omitted in the following.
The Wannier--Stark states $\Psi_n$ are orthogonal to their left eigenstates
$\Psi_m^L$ for $m \ne n$. Nevertheless, since we neglect the resonance
properties of the system we can identify left and right eigenvectors,
i.e. assume that $H_0$ is hermitian.
So multiplying equation (\ref{eqn-nlse-wsbasis})
by $\Psi_n^*$ and integrating yields
\be
  \ri \hbar \dot c_n = 2 \pi  F \, n \,  c_n + g \sum_{klm}
  \chi^n_{klm} c_k^* c_l c_m,
\ee
with the coupling tensor
\be
  \chi^n_{klm} = \int \Psi_n^*(x) \Psi_k^*(x) \Psi_l(x)
  \Psi_m(x) \, \rd x,
  \label{eqn-dnlse-def-chi}
\ee
which is symmetric under the exchange of its first and last two indices.
Due to the discrete translational invariance of the Wannier--Stark states
$\Psi_n(x) = \Psi_0(x-2 \pi n)$ one finds
\be
  \ri \hbar \dot c_n = 2 \pi  F \, n \,  c_n + g \sum_{klm}
  \chi_{klm} c_{k+n}^* c_{l+n} c_{m+n},
  \label{eqn-dnlse-full}
\ee
defining $\chi_{klm} \equiv \chi^0_{klm}$.

Though not suited for direct numerical calculations because of
the triple infinite sum, equation (\ref{eqn-dnlse-full}) provides
a basis for further approximations.
In the following we will reduce it to a simple integrable model, which
nevertheless captures important features of the dynamics.
To this end we decompose the coefficients $c_n$ into phase and amplitude
\be
  c_n = \sqrt{\rho_n} \, \re^{\ri \varphi_n}.
\ee
The imaginary parts of the coupling tensor $\chi_{klm}$ are negligible
and so one arrives at the coupled equations
\begin{eqnarray}
  \hbar \dot \varphi_n &=& - 2\pi F n - g \rho_n \sum_{klm} \chi_{klm}
  \left( \frac{\rho_{k+n} \rho_{l+n} \rho_{m+n}}{\rho_n^3}\right)^{1/2}
  \nonumber \\
  && \quad \quad  \times \cos(\varphi_{l+n}
  + \varphi_{m+n}   - \varphi_{k+n} - \varphi_n),   \label{eqn-dnlse-amp_phase1}
  \\[2mm]
   \hbar \dot \rho_n &=&\ 2 g \rho_n^2 \sum_{klm} \chi_{klm}
  \left( \frac{\rho_{k+n} \rho_{l+n} \rho_{m+n}}{\rho_n^3} \right)^{1/2}
  \nonumber \\
  && \quad \quad\times \sin(\varphi_{l+n}
  + \varphi_{m+n}   - \varphi_{k+n} - \varphi_n).
  \label{eqn-dnlse-amp_phase2}
\end{eqnarray}
If the initial state is broad, populating about 20 wells,
the amplitudes $\rho_n(t = 0)$ are small. Because of $\dot \rho_n\sim \rho_n^2$,
this implies that the amplitudes $\rho_n$ change only slowly in time compared to the
phases $\varphi_n$ and can be assumed to be constant.

Furthermore we reduce the expression for $\dot \varphi_n$ to the most important
contributions.
Numerically examining the $\chi_{klm}$ shows that the dominating terms are
$\chi_{000}$, $\chi_{kk0} = \chi_{k0k}$ and $\chi_{0kk}$, which is not
unexpected considering equation (\ref{eqn-dnlse-def-chi}).
It can also be argued (and verified numerically) that the terms in equation
(\ref{eqn-dnlse-amp_phase1}) which have a nonzero argument of the cosine have
little importance, as their contributions average out.
This leaves the terms including  $\chi_{000}$ and $\chi_{kk0} = \chi_{k0k}$
and finally leads to:
\begin{eqnarray}
  \hbar \dot \varphi_n &=& -2 \pi F n - g \gamma_n \rho_n
  \label{eqn-dnlse_lowest_approx}
\end{eqnarray}
with
\be
  \gamma_n = \chi_{000} + 2 \sum_{k\ne 0} \chi_{k0k}
  \frac{\rho_{n+k}}{\rho_n}.
  \label{eqn-estimate-alpha}
\ee
These equations (\ref{eqn-dnlse_lowest_approx}) are integrated to
\be
  \rho_n(t) = \rho_n(0)\ , \quad
  \varphi_n(t) = \omega_n t
  \label{eqn-dnlse_lowest_approx_sol}
\ee
with
\be
  \hbar \omega_n = - 2\pi F n - g\gamma_n \rho_n.
  \label{eqn-freq}
\ee
Note that this solution is exact for $g=0$.
Numerical calculations show that one can safely neglect the
dependence of  $\gamma_n$ on the index $n$ and set $\gamma_n \approx \gamma$.
For the given parameters equation (\ref{eqn-estimate-alpha}) yields
$\gamma_n \ge \gamma_0 = 0.278$. However, the best
fit with the results from a wave packet propagation are obtained for
$\gamma = 0.15$.

\begin{figure}[htb]
\centering
\includegraphics[width=8cm,  angle=0]{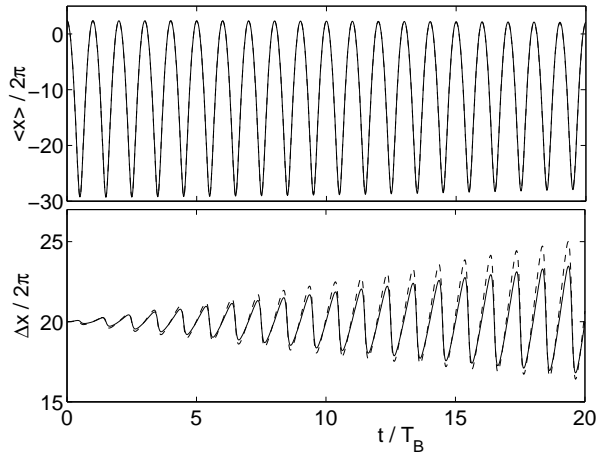}
\caption{\label{fig-so-dnlse-g=1}
Bloch Oscillations:
Expectation value of position $\langle x\rangle_t$ of the
wave packet for a moderate nonlinearity $g=1$.
The propagation was done with the split--operator method (dashed line) resp.~with approximation
(\ref{eqn-dnlse_lowest_approx_sol}) (solid line).}
\end{figure}

\begin{figure}[htb]
\centering
\includegraphics[width=8cm,  angle=0]{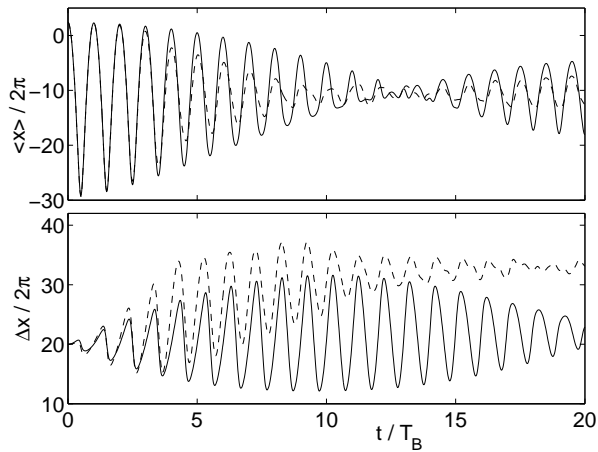}
\caption{\label{fig-so-dnlse-g=10}
As figure \ref{fig-so-dnlse-g=1}, but for a strong nonlinearity $g=10$.}
\end{figure}

This admittedly quite crude approximation shows very good
agreement with an exact numerical solution. In
figures \ref{fig-so-dnlse-g=1} and \ref{fig-so-dnlse-g=10}
the approximation (\ref{eqn-dnlse_lowest_approx_sol}) is compared with the
results obtained by a wave packet propagation using the split--operator
method \cite{Feit82}.
A normalized Gaussian initial distribution with coefficients
\be
  c_n \sim \re^{-n^2/4\sigma^2}, \quad \sigma =\tilde\sigma/2\pi= 20,
  \label{eqn-anfangszustand-basis}
\ee
is used which closely resembles a Gaussian wave packet projected onto
the lowest Bloch band in configuration space.
The dynamics for a moderate nonlinearity $g=1$ is well described by
equation (\ref{eqn-dnlse_lowest_approx_sol}), only the growth of the
width is somewhat underestimated. For $g=10$ the approximation
(\ref{eqn-dnlse_lowest_approx_sol}) becomes less accurate.
In particular it overestimates the revival of the
Bloch oscillation and underestimates the growth of the width of the wave
packet. However it still captures the important features at least
qualitatively: the decay and revival of the oscillations, the phase
jump around $t= 14 \,T_B$ and the breathing of the wave function.

\begin{figure}[htb]
\centering
\includegraphics[width=8cm,  angle=0]{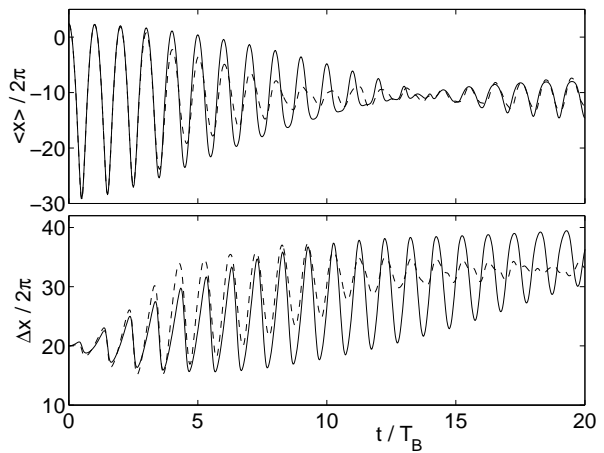}
\caption{\label{fig-so-dnlse-rvar-g=10}
Bloch Oscillations for a strong nonlinearity $g=10$.
The propagation was done with split--operator method (dashed line) resp.~with
approximation (\ref{eqn-rhodot-approx}) (solid line).}
\end{figure}

The systematic growth of the width of the wave packet is mainly due to a
broadening of the amplitude distribution $\rho_n$ and therefore clearly
not included in approximation (\ref{eqn-dnlse_lowest_approx_sol}).
To discuss this effect we briefly reintroduce the time--dependence
of the $\rho_n$.
Again we reduce the triple sum to keep the calculations feasible.
Note that the terms are oscillating due to the sine.
Using approximation (\ref{eqn-dnlse_lowest_approx_sol}) we see that
for $k=l+m$ the terms proportional to $F$ in the argument of the sine
cancel and hence the sine oscillates slowest. Thus we approximate
the dynamics of the amplitudes $\rho_n$ by
\begin{eqnarray}
  \hbar \dot \rho_n &\approx& 2 g \rho_n^2 \sum_{l,m} \chi_{l+m,l,m}
  \left( \frac{\rho_{l+m+n} \rho_{l+n} \rho_{m+n}}{\rho_n^3} \right)^{1/2}
  \nonumber \\
  && \quad \times \sin(\varphi_{l+n}
  + \varphi_{m+n}   - \varphi_{l+m+n} - \varphi_n),
  \label{eqn-rhodot-approx}
\end{eqnarray}
where the sum can be truncated at $|l|,|m| = 30$.
Equation (\ref{eqn-rhodot-approx})
for $\rho$ and equation (\ref{eqn-dnlse_lowest_approx}) for $\varphi$
are now solved numerically with $\gamma = 0.15$ and the initial condition (\ref{eqn-anfangszustand-basis}).
The results displayed in figure \ref{fig-so-dnlse-rvar-g=10} show
that this model captures the growth of the width of the wave packet.
We will, however, not go into details here and return to the
approximation (\ref{eqn-dnlse_lowest_approx_sol}) to discuss
the dynamics of Bloch oscillations.

\section{Analysis of the dynamical behavior}
\label{sec-koeffizienten-dynamik}

Further insight can be provided by a closer look at the dynamics of the
wave function in momentum space.
This can be achieved with the approximate time evolution of
the expansion coefficients $c_n$ derived in the previous section.

First we briefly reconsider the linear case.
For $g =0$ the equations (\ref{eqn-dnlse_lowest_approx_sol})
reduce to:
\be
  \rho_n(t) = \rho_n(t=0) \quad \mbox{and} \quad
  \varphi_n(t) = -2\pi F n t /\hbar.
  \label{eqn-linear-amp-phase}
\ee
The Wannier--Stark functions \cite{remark04} $\Psi_n$ are related by a spatial
translation $\Psi_n(x) = \Psi_0(x-2\pi n)$. In momentum space
this reads
\be
  \Psi_n(k) = \re^{-\ri 2\pi n k} \Psi_0(k)
\ee
and the time evolution of the wave function in momentum space is
\begin{eqnarray}
   \psi(k,t) &=& \Psi_0(k) \sum_n \sqrt{\rho_n} \exp \left(-\ri 2\pi n
   \left(k + Ft/\hbar \right) \right) \nonumber \\
   &\sim & \Psi_0(k) \, \widetilde C(k+Ft/\hbar),
   \label{eqn-bos-momentum}
\end{eqnarray}
neglecting a global phase. Thus the wave function is the product
of a time--independent function $\Psi_0(k)$ and the discrete Fourier
transformation $\widetilde C(k)$ of the amplitudes $\sqrt{\rho_n}$,
evaluated at the point $k+Ft/\hbar$.

The function $\widetilde C(k)$ is periodic in momentum space:
$\widetilde C(k+n) = \widetilde C(k)$ for $n \in \mathbb{Z}$.
Thus the function $\widetilde C(k+Ft/\hbar)$ is periodic in time
with the Bloch period $T_B = \hbar/F$.
For a broad Gaussian distribution of the amplitudes $\rho_n$
the discrete Fourier transform  $\widetilde C(k)$ is a
comb function with narrow peaks at $k = n$.

\begin{figure}[htb]
\centering
\includegraphics[width=7cm,  angle=0]{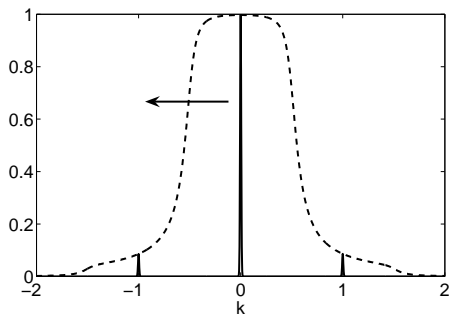}
\caption{\label{fig-bo_momentum_lin}
Bloch oscillations in momentum space. The wave function
$|\psi(k,t)|$ shown for $t=0$ (solid line) moves uniformly under
the envelope of the Wannier--Stark function $|\Psi_0(k)|$
(dashed line).}
\end{figure}

So one arrives at a simple view of the dynamics in momentum space:
The comb function $\widetilde C(k)$ moves uniformly under the envelope
$\Psi_0(k)$, as illustrated in figure \ref{fig-bo_momentum_lin}.
In coordinate space this periodic motion appears
as a Bloch oscillation \cite{04bloch1d}.

In the nonlinear case one has to evaluate
\begin{eqnarray}
   \psi(k,t) &=& \Psi_0(k) \sum_n \sqrt{\rho_n} \exp
   \left(-\ri \left( 2\pi n k - \varphi_n(t) \right) \right) \nonumber \\
   &=& \Psi_0(k) \, \widetilde C(k,t).
   \label{eqn-bos-momentum-nonlin}
\end{eqnarray}
instead of equation (\ref{eqn-bos-momentum}).
The time evolution of the phases is approximated according to equation
(\ref{eqn-dnlse_lowest_approx_sol}):
\be
  \varphi_n(t) = \omega_n t \quad \mbox{with} \quad
  \hbar \omega_n = - 2\pi F n - g \gamma \rho_n ,
  \label{eqn-lowest_approx_and_freq}
\ee
and the amplitudes are assumed to be Gaussian
\be
  \rho_n(t) = \rho_n \sim \re^{-n^2/2\sigma^2} \, .
\ee
As in the linear case, the static field term $- 2\pi F n$ in equation
(\ref{eqn-lowest_approx_and_freq}) for the frequency
leads to a uniform motion of the function $\widetilde C(k,t)$
in momentum space.
The nonlinear term $-g \gamma \rho_n$ leads to a dephasing of the coefficients
$c_n$ and broadens the Fourier transform $\widetilde C(k,t)$.
This dephasing causes a damping of the Bloch oscillations in coordinate space.

\begin{figure}[htb]
\centering
\includegraphics[width=8.5cm,  angle=0]{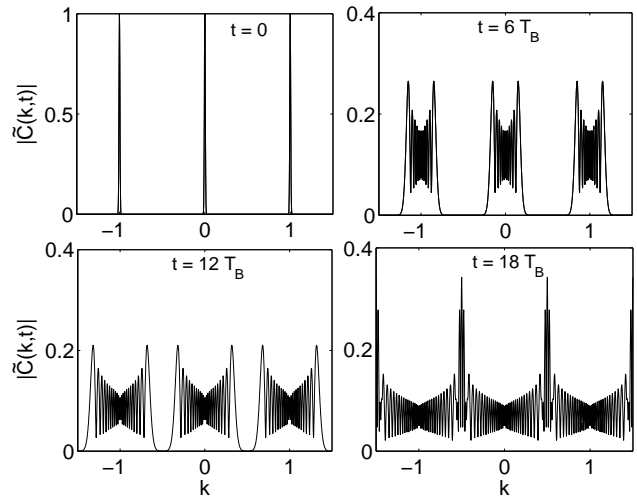}
\caption{\label{fig-dnlse-ctilde}
Time evolution of the function $\widetilde C(k,t)$ in equation
(\ref{eqn-bos-momentum-nonlin}) for $g=10$.
The function $\widetilde C$ is scaled as
$\widetilde C(0,0)=1$.}
\end{figure}

The oscillations of the width $\Delta x_t$, the breathing, can also
be understood with this approach.
In the linear case such breathing occurs for wave functions that
are initially strongly localized in coordinate space and thus have a broad
momentum distribution.
As explained, the nonlinear term leads to a broadening of
the function $\widetilde C(k,t)$ and hence to a broadening
of the wave function in momentum space and breathing in coordinate
space. For even longer times the coefficients dephase completely
and the oscillations in both the position and the width are
damped.

In order to understand the revivals and the phase jumps in the
oscillations of $\langle x \rangle_t$, we need
to look at the time evolution of the function $\widetilde C(k,t)$.
In figure \ref{fig-dnlse-ctilde} the function $|\widetilde C(k,t)|$
for $g=10$ is plotted at times $t=0$, $t = 6 \,T_B$, $t = 12 \,T_B$
and $t = 18 \, T_B$.
The dynamics of the expansion coefficients was calculated with
equation (\ref{eqn-dnlse_lowest_approx_sol}) resp.
(\ref{eqn-lowest_approx_and_freq}).
One observes that the initially narrow peaks are broadened and an
oscillatory structure develops with two maxima at the edges
of the band populated by $|\widetilde C(k,t)|$.
These maxima eventually merge, leading to a revival of the Bloch oscillations.
The new maximum after the merger is displaced by $\Delta k = 0.5$ in comparison
to the linear case and hence the phase of the Bloch oscillations is reversed.
This maximum broadens again, leading to a periodic breakdown and revival.
The phase of the Bloch oscillations is reversed after each breakdown
and the amplitudes of the revivals decrease.
However, these further revivals are observed only within the approximation
(\ref{eqn-dnlse_lowest_approx_sol}) and {\it not} in a wave packet propagation.

Now we consider the time dependence of the expectation values of
position and width.
These quantities can be evaluated analytically in the linear case
$g=0$ using a tight--binding approximation \cite{03TBalg,04bloch1d}.
In this approximation the expectation value of the position oscillates
harmonically with the Bloch frequency $\omega_B$,
\be
  \langle x\rangle_t=\bar x+A\cos (\omega_Bt)
  \label{eqn-bo-tb-amplitude1}
\ee
and amplitude
\be
  A=\frac{\Delta}{2 F} \, \re^{- 2\pi^2 \Delta p^2 / \hbar^2} \, ,
  \label{eqn-bo-tb-amplitude2}
\ee
where $\Delta p$ is the width in momentum space and $\Delta$ is
the bandwidth of the dispersion relation $E(\kappa)$ in the
field--free case.
For a small nonlinearity the broadening of the wave function in momentum
space due to the nonlinearity happens slowly compared to the
Bloch oscillations.
Thus we can assume that $\langle x \rangle_t$ still executes
damped harmonic oscillations with the amplitude (\ref{eqn-bo-tb-amplitude2}), where
the damping is determined by the slowly increasing momentum
width $\Delta p_t$.

According to equation (\ref{eqn-bos-momentum-nonlin}) we can estimate
the momentum width $\Delta p_t$  by the width of the peaks of the
function $\widetilde C(k,t)$.
For a broad distribution of the coefficients $c_n$, as assumed
throughout this paper, the sum in equation (\ref{eqn-bos-momentum-nonlin})
can be replaced by an integral:
\be
  \widetilde C(k,t) = \int_{-\infty}^{+\infty} \sqrt{\rho_n} \exp\left(-\ri (2 \pi n k -
  \varphi_n(t))\right) \, \rd n.
  \label{eqn-ctilde-integral}
\ee
This expression is valid for $|k| < 0.5$, otherwise $\widetilde C(k,t)$
is determined by its periodicity.
The amplitudes $\rho_n$ and phases $\varphi_n(t)$ are approximated by equation
(\ref{eqn-dnlse_lowest_approx_sol}), where the amplitudes $\rho_n$ are normalized as
\be
  \rho_n = \frac{1}{\sqrt{2\pi \sigma^2}} \,
  \re^{-n^2/2\sigma^2}.
\ee
We note that $\widetilde C(k,t)$ depends on the momentum $k$ only
through the expression $\tilde k = k + F t / \hbar$, reflecting the
uniform motion in momentum space due to the static field:
\be
  \widetilde C(\tilde k)\! = \! \frac{1}{\sqrt[4]{2\pi \sigma^2}}
  \!\int_{-\infty}^{+\infty}\! \! \!\re^{-n^2/4\sigma^2}\,
  \re^{-\ri \big(2\pi \tilde k n +\beta \re^{-n^2/2\sigma^2}\big)} \rd n
  \label{eqn-ctilde-integral-app}
\ee
with $\beta = \gamma g  t / (\sqrt{2\pi} \sigma \hbar)$.

\begin{figure}[htb]
\centering
\includegraphics[width=7cm,  angle=0]{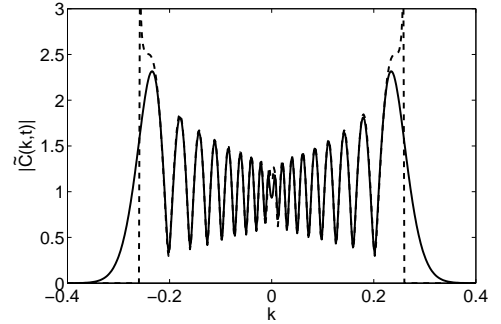}
\caption{\label{fig-ctilde_compare}
Function $|\widetilde C(k,t)|$ (\ref{eqn-ctilde-integral}) for
$g t = 90 \, T_B$ and $\gamma = 0.15$. The integral was evaluated
numerically (solid line) resp. using the stationary phase method
(dashed line).}
\end{figure}

The integral (\ref{eqn-ctilde-integral}) can be evaluated using the
stationary phase approximation. However, there exists only a finite
$\tilde k$--interval for which stationary points exist.
For $|\tilde k| > |k_c|$ with
\be
  k_c = \frac{\beta}{2 \pi \sigma \re^{1/2}}
  \label{eqn-critical-k}
\ee
the integral vanishes in the simple stationary phase approximation \cite{remark05}.
For $|\tilde k| < |k_c|$ one obtains
\be
  \widetilde C(k,t) \approx  \left({\frac{2\pi\sigma^2}{\beta^2}}\right)^{1/4}
    \left( \frac{\re^{\ri \zeta_+}}{\sqrt{1 - z_+}} +
    \frac{\re^{\ri \zeta_-}}{\sqrt{1 - z_-}} \right),
  \label{eqn-ctilde_statphase}
\ee
where $z_\pm$ are the two solutions of the equation
\be
  z \re^{-z} = \left({2 \pi \sigma \tilde k}/{\beta}\right)^2
\ee
and the abbreviations $\zeta_\pm = - 2 \pi \sigma \tilde k \,
(z_\pm^{+1/2}+z_\pm^{-1/2})$ were used.
As an example the function $|\widetilde C(k,t)|$ is plotted in
figure \ref{fig-ctilde_compare} for $g t = 90 \, T_B$ and
$\gamma=0.15$.

Estimating the momentum width as $\Delta p \approx \hbar |k_c|$
one arrives at
\be
  \Delta p_t \approx  \frac{|\gamma g|}{(2 \pi)^{3/2} \sigma^2\re^{1/2}} \, t.
\ee
Thus the damped Bloch oscillations in coordinate space are described by
\be
  \langle x \rangle_t \approx \bar x +
  \frac{\Delta}{2 F} \exp\left(- \frac{\gamma^2 g^2 t^2}{
  4 \pi \re \hbar^2 \sigma^4}\right) \,
    \cos(\omega_B t)
  \label{eqn-tbmodel-xev}
\ee
according to (\ref{eqn-bo-tb-amplitude1}).
The amplitude decreases exponentially with $-g^2 t^2$
in agreement with the estimate given in \cite{Trom01}.

Furthermore we can calculate approximately the time up to the first rephasing of the
coefficients and thus the first revival of the Bloch oscillations.
This revival occurs if the outer peaks of $\widetilde C(k,t)$ meet
at $k = n + 1/2, \, n \in \mathbb{Z}$, as illustrated in figure
\ref{fig-dnlse-ctilde}. Therefore the first rephasing and revival
occurs for $k_c = 0.5$ which yields a revival time
\be
   t_{\rm rev} \approx \frac{(2 \pi)^{3/2} \re^{1/2} \, \hbar
  \sigma^2}{2 |\gamma g|}\,.
\ee
For $g= 10$ and $\gamma = 0.15$ one obtains
\be
  t_{\rm rev} \approx 17 \, T_B
\ee
in reasonable agreement with the revival of Bloch oscillations
observed numerically for the wave packet propagation
shown in figure \ref{fig-bonl_long_xev}.

For very long times the coefficients $c_n$ dephase completely.
We can therefore estimate the position expectation value by approximating
the wave function as an incoherent sum of the basis states.
Assuming that the amplitudes $\rho_n$ are constant in time according
to equation (\ref{eqn-dnlse_lowest_approx}) one has
\be
  \langle x \rangle = \sum_n \rho_n \langle \Psi_n|x| \Psi_n \rangle.
\ee
Using the translational properties of the Wannier--Stark
states (cf. \cite{02wsrep}) one arrives at
\be
  \langle x \rangle_{\infty} \approx \langle \Psi_0 |x| \Psi_0 \rangle
  + 2\pi \, \sum_n n \rho_n.
\ee
The amplitudes of the initial state (\ref{eqn-anfangszustand-basis})
are symmetric around $n=0$ and hence this approximation yields
$\langle x \rangle_{\infty} \approx \langle \Psi_0 |x| \Psi_0 \rangle
=-10.5 \cdot 2\pi$. This estimation fairly agrees with the numerical
results displayed in figure \ref{fig-bonl_long_xev}.
As argued above, the systematic growth of the width of the wave packet
is mainly due to a broadening of the amplitude distribution $\rho_n$ and hence
cannot be explained using the simple model discussed here.
\section{Strong Static field and decay}
\label{sec-basis-hohe-felder}

An expansion into Wannier--Stark resonances is also very helpful
in order to understand the dynamics and decay in strong static fields.
In the following we will discuss the dynamics for the parameters
$\hbar = 3.3806$ and $F = 0.0661$, corresponding to the experiment
of the Kasevich group \cite{Ande98}.
A detailed discussion of this experiment in terms of Wannier--Stark
resonances, however neglecting the nonlinearity, can be found in
\cite{02pulse}. Thus we will only briefly discuss the influence of
the nonlinearity on the pulse shape.

For a field strength of $F = 0.0661$ decay cannot be neglected
any longer. One has to take into account that the resonance states
eventually diverge exponentially for $x$ or $k \rightarrow - \infty$.
Hence, a wave function of the form (\ref{eqn-bos-momentum-nonlin})
is not normalizable.
Nevertheless, the restriction to the ground Wannier--Stark
ladder is still sufficient.

As described in \cite{02pulse} (see also remark \cite{remark04})
one can solve the problem of
normalization by introducing truncated resonance states defined by
\be
   \Psi_n^K (k) = \Theta(k+K) \Psi_n(k).
\ee
The Heaviside--function $\Theta(k+K)$ truncates the
resonances at $-K$. Provided that $|K|$ is large enough, the
time evolution of these states is given by
\be
  \Psi_n^K (k,t) = \Theta(k+K+Ft/\hbar) \Psi_n(k,t).
\ee
If the support of the initial wave function is bounded
in momentum space by $|k| < |K|$, we can expand it
into a basis of truncated resonances.
The dynamics of this state is then given by
\be
   \psi(k,t) =
   \Theta(k + K + Ft/\hbar) \, \Psi_0(k,t) \, \widetilde C(k,t),
   \label{eqn-pulsedoutput-momentum-nonlin}
\ee
with $\Psi_0(k,t) = \exp(-\ri {\cal E}_0 t /\hbar) \Psi_0(k)$
instead of equation (\ref{eqn-bos-momentum-nonlin}).

For a coherent initial distribution of a sufficient width
$c_n \sim \exp(-n^2/(2 \sigma)^2)$ with $\sigma \gg1$, the function
$\widetilde C(k,t)$ is a comb function
in the linear case, leading to a pulsed output.
The pulse shape given by the function $\widetilde C$ broadens and
deforms under the influence of the nonlinearity as described in the
previous section (cf. figure \ref{fig-dnlse-ctilde}).
This deformation is directly observable in the pulsed output.

\begin{figure}[thb]
\centering
\includegraphics[width=8.5cm,  angle=0]{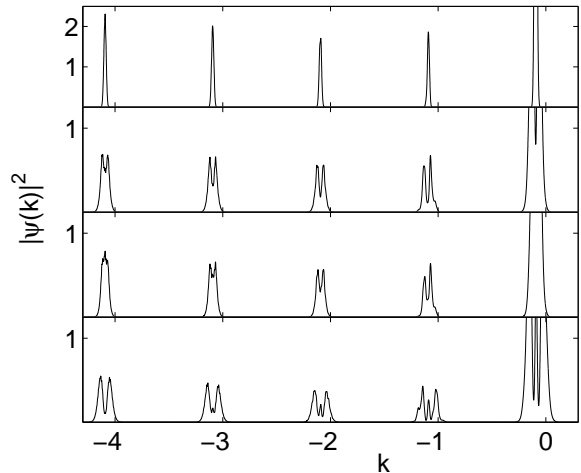}
\caption{\label{fig-pulse_shape}
Pulsed output for different nonlinearities $g=0,\, -5,\, 5$
and $10$ (from top to bottom). The wave function
$|\psi(k,t)|^2$ is displayed for $t = 9.1 \, T_B$.}
\end{figure}

This is demonstrated in figure \ref{fig-pulse_shape} for a coherent
initial distribution $c_n \sim \exp(-n^2/4 \sigma^2)$ with a width
of $\sigma = 15/2$. The time evolution was again calculated using
a split--operator method.
The wave function $|\psi(k,t)|^2$ is plotted at a time $t = 9.1 \, T_B$
for four different values of the nonlinearity  $g=0, \, -5, \, 5$
and $10$.
The broadening of the peaks with increasing $|g|$  and the
characteristic double--peak structure can clearly be seen.

The resulting wave function in coordinate space is a sequence of pulses
at the points
\be
  x = x_0 - \frac{F}{2} ( t + j \, T_B)^2,
\ee
with $x_0 = E_0/F$ and $j \in \mathbb{Z}$. These pulses are accelerated
just like classical particles in a static field, as observed in the
experiment \cite{Ande98}. The pulse shape is approximately described by
the discrete Fourier transform of $\widetilde C$ \cite{02pulse}.
Thus one also finds a characteristic deformation of the pulses in
coordinate space.

\section{Conclusions}
In this article we first investigated Bloch oscillations of BECs by numerical
solutions of the Gross--Pitaevskii equation (GPE) and demonstrated a revival
of Bloch oscillations after an initial breakdown. These findings have been
further analyzed via discretising the GPE
in a Wannier--Stark basis set expansion. Using these resonance states
one can easily compare the linear and nonlinear case. This comparison leads
to a better understanding of the nonlinear features of BECs in optical lattices.
It allows us to derive a simple integrable model (\ref{eqn-dnlse_lowest_approx_sol})
which can explain the nonlinear phenomena of breakdown and revival of the Bloch
oscillations.
This approach, unlike the tight--binding approximation, works as well for
strong Stark fields.

Many interesting questions are left open and deserve future studies,
as for example the following: (a) The effects induced by the nonlinearity for
Bloch oscillations in two--dimensional lattices, where recently
novel effects concerning the extreme sensitivity on the field direction
with respect to the lattice have been found \cite{04bloch2d,04tb2d}.
(b) For BECs in tilted optical lattices a
classical chaotic behavior has been reported \cite{Thom03}. These
interesting findings deserve further studies, for example the
correspondence of the emergence of chaos with the loss of stability
of the GPE solutions.
(c) It is an entirely open question how the nonlinearity influences
the behavior of a driven Wannier--Stark system, e.g. the stabilizing
phenomena found for an additional harmonic driving
\cite{00transition2,02wsrep}.

%
\begin{acknowledgments}
Support from the Deutsche Forschungsgemeinschaft
as well as from the Volkswagen Foundation is gratefully acknowledged.
We thank Jean--Claude Garreau, Andrey Kolovsky and Andreas Buchleitner for providing results
prior to publication.
\end{acknowledgments}


\begin{thebibliography}{10}
\bibitem{remark04}
Though not important for weak fields, it should be noted that the
  Wannier--Stark states are not normalizable because of their exponentially
  increasing tail at $x\to -\infty$. In practice, this can be circumvented by a
  cutoff in coordinate space or momentum space. For the stronger fields
  considered in section \ref{sec-basis-hohe-felder} this truncation must be
  explicitly taken into account.
\bibitem{remark05}
If desired, the stationary phase points can be continued into the complex plane
leading to an exponential decay.
\end{thebibliography}

\end{document}